\documentclass[12pt]{article}
\usepackage{jt-article}
\usepackage{jt-briefnote}
\usepackage{jt-bea}
\usepackage{psfig}
\date{October 19, 1998}

\providecommand{\bi}{\begin{itemize}}
\providecommand{\ei}{\end{itemize}}
\newcommand{\nsrc}{N}
\newcommand{\nbak}{B}
\newcommand{\asrc}{A_{\mbox{\footnotesize src}}}
\newcommand{\abak}{A_{\mbox{\footnotesize bak}}}

\newcommand{\plamp}{p_{\mbox{\scriptsize Lamp}}}
\newcommand{\palex}{p_{\mbox{\scriptsize Alex}}}

\newcommand{\binom}[2]{\left(\begin{array}{cc} #1 \\ #2 \end{array}\right)}

\begin{document}

\briefnote{A few brief notes\ldots}
\title{On the Equivalence of Two Expressions\\
	for Statistical Significance in Point Source Detections}
\author{James Theiler\\
        {\tt jt@lanl.gov}}
\address{Space and Remote Sensing Sciences Group, NIS-2\\
        Los Alamos National Laboratory, Los Alamos, NM 87545}
\maketitle

\begin{abstract}

The problem of point source detection in Poisson-limited count maps
has been addressed by two recent papers [M.~Lampton, {\em ApJ} {\bf
436}, 784 (1994); D.~E. Alexandreas, {\it et al.}, {\em
Nucl. Instr. Meth. Phys. Res. A} {\bf 328}, 570 (1993)].  Both papers
consider the problem of determining whether there are significantly
more counts in a source region than would be expected given the number
of counts observed in a background region.  The arguments in the two
papers are quite different (one takes a Bayesian point of view and the
other does not), and the suggested formulas for computing $p$-values
appear to be different as well.  It is shown here that
the expressions provided by the authors of these two articles are in
fact equivalent.
\end{abstract}

\section{Introduction}

Space is big.  Stars are big too, but space is bigger.  As a
consequence, stars and in fact most astronomical objects of interest
appear as point sources from our perspective here on earth.  Thus many
astronomical surveys concentrate on the detection and characterization
of point sources in the sky.  Particularly for instruments that
measure high energy radiation (extreme ultraviolet and beyond),
individual photons are counted, and for these instruments
the statistical treatment of point
source detection requires proper consideration of Poisson statistics.

The particular problem of interest is the following: {\it Given $N$
counts in a source region of area $\asrc$, and $B$ counts in a
background region of area $\abak$, is there a real point source in the
source kernel?}  More specifically, compute a $p$-value associated
with the probability of observing $N$ or more counts in the source
region under the null hypothesis that the count rate per unit area is
the same for both the source and background regions.  In two recent
papers~\cite{Lampton94,Alexandreas93}, this specific problem of point
source detection is addressed.  The papers take quite different
approaches in their derivation, and produce expressions which appear
on the surface to be quite different.  It will be shown, however, that the
expressions are equal.

\subsection{Binomial formulation}

Lampton~\cite{Lampton94}, hereafter referred to as Paper~I, provides an 
elegant formulation.
Rather than consider the source and background as separate Poisson
processes, the sum $\nsrc+\nbak$ is treated as a fixed number,
and the binomial distribution of $\nsrc+\nbak$ total counts into areas $\asrc$
and $\abak$ is considered.  Let $f=\asrc/(\asrc+\abak)$ be the fraction
of total counts expected in the source region; then $1-f$ is the
fraction of counts expected in the background region.
We can write down the likelihood that exactly $n$ counts would be observed 
in the source region:
\be
	P(n) = \binom{\nsrc+\nbak}{n} f^n(1-f)^{\nsrc+\nbak-n}
\ee
where 
\be
	\binom{a}{b} = \left\{\begin{array}{cl}
	\displaystyle\frac{a!}{b!(a-b)!} & \quad\mbox{for $a\ge b\ge 0$} \\
	0 & \quad\mbox{otherwise}\end{array}\right.
\ee
is the binomial coefficient.  The $p$-value
is given by the probability of observing 
$n\ge\nsrc$ photons in the source region:
\be
	\plamp = \sum_{n=\nsrc}^{\nsrc+\nbak}
	\binom{\nsrc+\nbak}{n}f^n(1-f)^{\nsrc+\nbak-n}
	\label{eq:Lampton}
\ee

\subsection{Bayesian formulation}

In Alexandreas~{\it et~al.}~\cite{Alexandreas93}, 
hereafter referred to as Paper~II, the argument
begins with the remark that if
the expected number $\mu$ of counts in the source region were known
exactly, then likelihood of seeing exactly
$n$ counts in the source region is given by the Poisson formula:
\be
	P(n|\mu)=\frac{\mu^n}{n!}e^{-\mu}.
\ee
The $p$-value is the probability of observing $n\ge\nsrc$ counts:
\be
	p = \sum_{n=\nsrc}^\infty \frac{\mu^n}{n!}e^{-\mu}.
	\label{eq:Poisson-pvalue}
\ee
In the problem at hand, the actual background level is not known exactly,
but it can be estimated from the $\nbak$ counts in 
the background region.  Following the notation in Paper~II, let 
$\alpha=\asrc/\abak$ be the ratio of areas in the source and background 
region.  Then $\hat\mu = \alpha \nbak$ is an estimate for the expected
number of counts in the source region.  In Paper~II, the authors 
implicitly take a Bayesian approach (though they do not identify it as 
Bayesian) and produce a probability distribution on the parameter $\mu$ 
that is proportional to the likelihood of observing $\nbak$ background
counts, given $\mu$.
\be
	P(\mu|\nbak) = \frac{P(\mu)P(B|\mu)}{P(B)}
	\propto P(\nbak|\mu) = 
	\frac{(\mu/\alpha)^{\nbak}}{\nbak!}
	e^{-\mu/\alpha}
\ee
This direct proportionality implies that a uniform Bayesian prior was used
({\it i.e.,} $P(\mu)=\mbox{constant}$);
since $\mu$ is unbounded, this is a so-called ``improper'' prior.
(Loredo~\cite{Loredo90} has suggested that the prior $P(\mu)=1/\mu$
is more appropriate.)
Based on this distribution, 
an ``average'' $p$-value is computed, by integrating
the expression for $p$ in \eq{Poisson-pvalue} against $P(\mu|\nbak)$.
That is,
\be	
	p = \sum_{n=\nsrc}^\infty \frac{1}{n!}\int_0^\infty \mu^n e^{-\mu}
	\frac{(\mu/\alpha)^{\nbak}}{\nbak!}
	e^{-\mu/\alpha}\,d\mu
\ee
After performing the integral, a closed form series solution is obtained,
and the form 
of this expression given in Paper~II is
\be
	\palex = 1 - \sum_{n=0}^{\nsrc-1}
	\frac{\alpha^{n}}{(1+\alpha)^{(n+\nbak+1)}}
	\,\frac{(n+\nbak)!}{n!\nbak!}.
	\label{eq:Alexandreas}
\ee
To facilitate comparison with the formula in Paper~I, we will use
$\alpha=f/(1-f)$, employ the standard notation for binomial coefficients,
and use the fact that $\sum_{n=0}^{\infty}=1$ to write an equivalent
form:
\be
	\palex = (1-f)^{\nbak+1}\,
	\sum_{n=\nsrc}^{\infty} \binom{n+\nbak}{n}f^n
	\label{eq:Alex}
\ee

\section{Proof of Equivalence}

We will begin with three simple lemmas.

\vspace{2ex} %% Ta Da!

{\bf Lemma 1.} {\it The following is an identity:}
\be
	\binom{a}{b} + \binom{a}{b+1} = \binom{a+1}{b+1}.
	\label{eq:Lemma1}
\ee
This is in fact a well-known identity, but we will invoke it several times in 
the proofs below. 

%%  To see that it is true, write the binomials explicitly
%% as ratios of factorials:
%% \bea
%% 	\binom{a}{b} + \binom{a}{b+1} &=&
%% 	\frac{a!}{b!(a-b)!} + \frac{a!}{(b+1)!(a-b-1)!} 
%% 	\nonumber \\
%% 	&=& \frac{(b+1)a!}{(b+1)!(a-b)!} + \frac{(a-b)a!}{(b+1)!(a-b)!} 
%% 	\nonumber \\
%% 	&=& \frac{(b+1+a-b)a!}{(b+1)!(a-b)!} = \frac{(a+1)!}{(b+1)!(a-b)!} 
%% 	= \binom{a+1}{b+1}.
%% \eea

\vspace{2ex} %% Ta Da!

{\bf Lemma 2.} {\it The following expression holds for $n\ge0$ and $0\le f < 1$:}
\be
	\frac{f^n}{(1-f)^{n+1}} = \sum_{k=0}^\infty \binom{k}{n} f^k.
	\label{eq:Lemma2}
\ee

The proof of this is fairly straightforward.  The binomial theorem
for $(1-f)^{-n-1}$ produces the infinite series, and then both sides
are multiplied by $f^n$.  Note that the summation starts at $k=0$,
even though the first nonzero term is $k=n$.  

\vspace{2ex} %% Ta Da!

{\bf Lemma 3.} {\it The following is an identity:}
\be
	\binom{n+\nsrc+\nbak}{n+\nsrc} =
	\sum_{k=0}^{\nbak} \binom{\nsrc+\nbak}{\nsrc+k}\binom{n}{k}.
	\label{eq:Lemma3}
\ee

This is proved by induction on $n$.
First note that $n=0$ produces $\binom{\nsrc+\nbak}{\nsrc}$
on both the left and right sides, so the statement is true for $n=0$.
Now, suppose it is true for $n=n_o$, and consider the case $n_o+1$.
First expand out the right hand side, using Lemma 1:
\be
	\sum_k \binom{\nsrc+\nbak}{\nsrc+k}\binom{n_o+1}{k} =
	\sum_k \binom{\nsrc+\nbak}{\nsrc+k}\left[
	\binom{n_o}{k} + \binom{n_o}{k-1}\right].
\ee
Rearrange the terms
\be
	\sum_k \binom{\nsrc+\nbak}{\nsrc+k}\left[
	\binom{n_o}{k} + \binom{n_o}{k-1}\right] =
	\sum_k 
	\left[\binom{\nsrc+\nbak}{\nsrc+k} 
		+ \binom{\nsrc+\nbak}{\nsrc+k+1}\right]	
	\binom{n_o}{k}.
\ee
Again, apply Lemma 1:
\be
	\sum_k 
	\left[\binom{\nsrc+\nbak}{\nsrc+k} 
		+ \binom{\nsrc+\nbak}{\nsrc+k+1}\right]	
	\binom{n_o}{k}
	= 
	\sum_k \binom{\nsrc+\nbak+1}{\nsrc+k+1}\binom{n_o}{k}.
\ee
We have inductively assumed that \eq{Lemma3} is valid for $n=n_o$, so
\be
	\sum_k \binom{\nsrc+\nbak+1}{\nsrc+k+1}\binom{n_o}{k}
	= \binom{n_o+\nsrc+\nbak+1}{n_o+\nsrc+1}.
\ee
Combining all of these produces
\be
	\sum_k \binom{\nsrc+\nbak}{\nsrc+k}\binom{n_o+1}{k}
	= \binom{n_o+\nsrc+\nbak+1}{n_o+\nsrc+1},
\ee
which is \eq{Lemma3} for $n=n_o+1$.  Thus, our induction was successful, and 
Lemma 3 is proved.

\vspace{2ex} %% Ta Da!

{\bf Theorem.} {\it
The expressions for $p$-value in \eq{Alex} and 
in \eq{Lampton} are equivalent.  That is:}
\be
	\palex = \plamp.
\ee

To see this,
start with the expression for $p$-value in \eq{Alex}, and substitute
$k=n-\nsrc$:
\be
	\palex = (1-f)^{\nbak+1}\,
	\sum_{n=\nsrc}^{\infty} \binom{n+\nbak}{n}f^n
	= (1-f)^{\nbak+1}\,f^\nsrc
	\sum_{k=0}^{\infty} \binom{k+\nsrc+\nbak}{k+\nsrc}f^k
\ee
Use the identity in Lemma 3:
\bea
	\palex &=& (1-f)^{\nbak+1}\,f^\nsrc\,
	\sum_{k=0}^{\infty} \binom{k+\nsrc+\nbak}{k+\nsrc}f^k \\
	&=& 
	(1-f)^{\nbak+1}\,f^\nsrc\,
	\sum_{k=0}^{\infty} 
	\left[\sum_{j=0}^{\nbak} \binom{\nsrc+\nbak}{\nsrc+j}\binom{k}{j}
	\right]\,f^k.
\eea
Switch the order of summation
\be
	\palex = (1-f)^{\nbak+1}\,f^\nsrc\,
	\sum_{j=0}^{\nbak} \left[\binom{\nsrc+\nbak}{\nsrc+j}
	\sum_{k=0}^\infty \binom{k}{j}\,f^k\right].
\ee
Substitute the result from Lemma 2 in \eq{Lemma2}:
\bea
	\palex &=& (1-f)^{\nbak+1}\,f^\nsrc\,
	\sum_{j=0}^{\nbak} \binom{\nsrc+\nbak}{\nsrc+j}
	\frac{f^j}{(1-f)^{j+1}} \\
	&=& 
	\sum_{j=0}^{\nbak} \binom{\nsrc+\nbak}{\nsrc+j}
	f^{\nsrc+j}(1-f)^{\nbak-j}.
\eea
Finally, substitute $n=j+\nsrc$ to obtain
\bea
	\palex &=& 
	\sum_{n=\nsrc}^{\nsrc+\nbak} \binom{\nsrc+\nbak}{n}
	f^n(1-f)^{\nsrc+\nbak-n}\\
	&=& \plamp.
\eea
QED

\section{Discussion}

Though the same answer is ultimately obtained, this does not imply that the
two approaches are equivalent.  Hypothesis testing in the presence of a
nuisance parameter, in this case the background level, is always problematic.

In Paper~I, a trick is employed which enables us to express the null
hypothesis in terms that are independent of the nuisance parameter.
Crucial use is made of the identity which expresses the joint
distribution of two Poisson processes (the counts in the source and background
regions) as a product of a single Poisson process and a binomial process:
\bea
	{\cal P}(\nsrc,\mu)\,{\cal P}(\nbak,\mu/\alpha) &=&	
	\frac{\mu^\nsrc}{\nsrc!}e^{-\mu}\,\times\,
	\frac{(\mu/\alpha)^\nbak}{\nbak!}e^{-\mu/\alpha}
	\nonumber \\
	&=&
	\frac{(\mu+\mu/\alpha)^{\nsrc+\nbak}}{(\nsrc+\nbak)!}e^{-\mu-\mu/\alpha}
	\,\times\,\frac{(\nsrc+\nbak)!}{\nsrc!\nbak!}\,
	\frac{\mu^\nsrc(\mu/\alpha)^\nbak}{(\mu+\mu/\alpha)^{\nsrc+\nbak}} 
	\nonumber \\
	&=&
	{\cal P}(\nsrc+\nbak,\mu+\mu/\alpha)\,
	{\cal B}(\nsrc,\nbak,\frac{\alpha}{1+\alpha})
\eea
The single Poisson process describes the
statistics on the total $\nsrc+\nbak$, while the binomial process
describes the partition of this total into the source and background regions.
Since the joint distribution is a simple product, the two processes are
independent.
It is important to note that the binomial process does not depend on $\mu$;
thus, there is no need to worry about estimating this parameter in testing
the null hypothesis.

Such decompositions are not always available, particularly as the
problems get more complicated, but Paper~II provides a methodology
that is more adaptable to such situations: one estimates a distribution
for the nuisance parameter, and then integrates over it.  However, this
uses a Bayesian derivation to produce a fundamentally ``frequentist''
product, namely a $p$-value.  
%This is arguably dubious as a general method.  
Given that the choice of prior is arbitrary
(the choice in this case was quite natural, though Loredo~\cite{Loredo90}
has argued that other choices might be preferred), one might say that
the authors of Paper~II were ``lucky'' to get the same answer that was
obtained in Paper~I without any free choices.

Loredo~\cite{Loredo90} argues (quite strenuously!) for a purely Bayesian
approach, producing in the end a probability distribution
on a source strength parameter.  It is not clear how to compare this result
with the $p$-values produced by the methods~\cite{Lampton94,Alexandreas93}
discussed here.

\section*{Acknowledgement}

This work was conducted under the auspices of the United States Department
of Energy.

\bibliography{apjmnemonic,Alexis/alexis-rel-x}
\bibliographystyle{jt}

\end{document}